\documentclass[aps,prl,twocolumn,superscriptaddress]{revtex4}

\usepackage{amssymb}
\usepackage{graphicx}
\usepackage{dcolumn}
\usepackage{bm}
\usepackage{color}
\usepackage{ulem}

\begin{document}

\title{Optical spin transfer torque driven domain wall motion in ferromagnetic semiconductor}

\author{A.~J.~Ramsay}
\email{ar687@cam.ac.uk}
\affiliation{Hitachi Cambridge Laboratory, Hitachi Europe Ltd.,  Cambridge CB3 0HE, UK}

\author{P.~E.~Roy}
\affiliation{Hitachi Cambridge Laboratory, Hitachi Europe Ltd.,  Cambridge CB3 0HE, UK}

\author{J.~A.~Haigh}
\affiliation{Hitachi Cambridge Laboratory, Hitachi Europe Ltd.,  Cambridge CB3 0HE, UK}

\author{R.~M.~Otxoa}
\affiliation{Hitachi Cambridge Laboratory, Hitachi Europe Ltd.,  Cambridge CB3 0HE, UK}

\author{A.~C.~Irvine}
\affiliation{The Cavendish Laboratory, University of Cambridge,  Cambridge CB3 0HE, UK}

\author{T.~Janda}
\affiliation{Faculty of Mathematics and Physics, Charles University in Prague, Ke Karlovu 3, 121 16 Prague 2, Czech Republic}

\author{R.~P.~Campion}
\affiliation{School of Physics and Astronomy,University of Nottingham, Nottingham, NG7 2RD, UK}

\author{B.~L.~Gallagher}
\affiliation{School of Physics and Astronomy,University of Nottingham, Nottingham, NG7 2RD, UK}

\author{J.~Wunderlich}
\affiliation{Hitachi Cambridge Laboratory, Hitachi Europe Ltd.,  Cambridge CB3 0HE, UK}
\affiliation{Institute of Physics ASCR, v.v.i., Cukrovarnick\'{a} 10, 162 53 Praha 6, Czech Republic}

\date{\today}

\begin{abstract}
We demonstrate optical manipulation of the position of a domain wall in a dilute magnetic semiconductor, GaMnAsP. Two main contributions are identified. Firstly, photocarrier spin exerts a spin transfer torque on the magnetization via the exchange interaction. The direction of the domain wall motion can be controlled using the helicity of the laser. Secondly, the domain wall is attracted to the hot-spot generated by the focused laser.  Unlike magnetic field driven domain wall depinning, these mechanisms directly drive domain wall motion, providing an optical tweezer like ability to position and locally probe domain walls.

\end{abstract}

\maketitle



Recent work advocating the use of mobile magnetic domains for memory \cite{Parkin} and logic \cite{Allwood} applications has sparked renewed interest in the physics of domain wall motion. A key element in these research efforts is the controlled propagation of domain walls driven by electric currents \cite{berger,Freitas,Yamaguchi,Mougin,reviewSTT}. This occurs via spin transfer torque where, due to an exchange interaction with the magnetization, a non-collinear injected carrier spin can exert a torque on the magnetization. Recently, it has been demonstrated that a spin transfer torque can also be applied optically \cite{Nemec_nphys}.  In that experiment, the optical spin transfer torque  was used to induce precession \cite{Tesarova_nphoton,Hall_apl} of the magnetization in a thin-film of $\mathrm{GaMn_{<0.09}As}$ with an in-plane easy axis.

Here we consider a similar material, $\mathrm{Ga_{0.94}Mn_{0.06}As_{0.91}P_{0.09}}$, with an out-of-plane easy-axis. In this case, the optical spin transfer torque induced by a circularly polarized laser at normal incidence is not expected to act within the magnetic domain. However at a  domain boundary, the magnetization has an in-plane component, allowing a local torque that results in an optical polarization dependent domain wall (DW) motion. We demonstrate this helicity dependent optically induced DW motion by exposing a single DW to a train of above bandgap picosecond laser pulses, and identify the optical spin transfer torque as the dominant helicity dependent mechanism driving the DW motion. In addition, we observe a second helicity independent effect that attracts the DW to the center of the focused Gaussian laser spot due to local heating of the magnetic material.
During laser exposure the DW moves towards a final position where the effects of the optical spin transfer torque and the thermal gradient are balanced. This interpretation is confirmed by numerical simulations based on the Landau-Lifshitz-Bloch equations.  

The wafer consists of a 25-nm thick film of $\mathrm{Ga_{0.94}Mn_{0.06}As_{0.91}P_{0.09}}$ on a GaAs substrate. The annealed sample has a Curie Temperature of 106 K. The addition of P results in an out-of-plane easy-axis via tensile growth strain \cite{Rushforth}. To study domains constrained to one spatial dimension, the wafer is fabricated into $4\times 60~\mathrm{\mu m}$ bars. Further details of the sample can be found in ref. \cite{deRanieri_nmat}.

The sample is mounted in a cold-finger cryostat at 92~K. An  out-of-plane magnetic field can be applied using an electro-magnet. The magnetic domains are imaged using a  Kerr-microsope. A mode-locked Ti:sapphire laser provides a  source of 140-fs optical pulses at an 80-MHz repetition rate. A bar aligned along the $[1\bar{1}0]$ direction, with a Neel wall \cite{deRanieri_nmat}, is excited with  an exposure time of $>4~\mathrm{ms}$ using a mechanical shutter. After a 10-m single-mode fiber, dispersion stretches the pulses to approximately 4-ps. 
The laser is focused to a spot with a Gaussian intensity profile with a full width at half maximum of $w=5~\mathrm{\mu m}$.


\begin{figure}[ht!]
\begin{center}
\vspace{0.2 cm}
\includegraphics[width=\columnwidth]{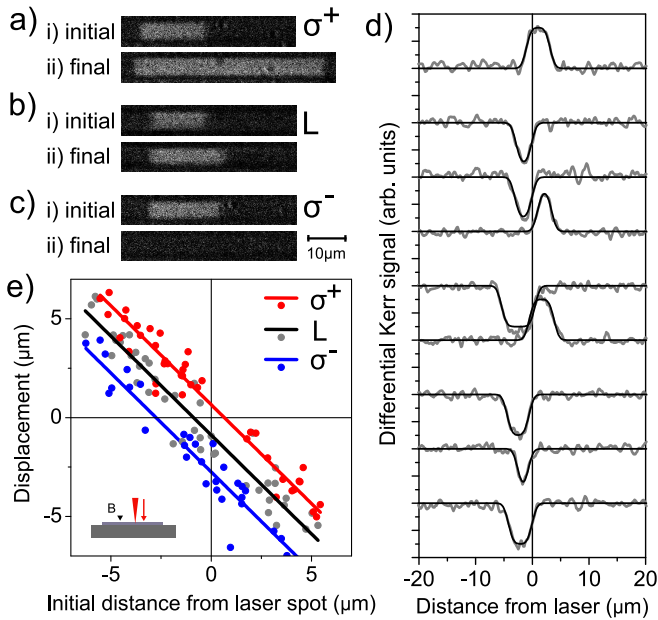}
\vspace{0.2 cm}
\end{center}
\caption{(a,b,c) Images of the initial domain nucleated by the nucleation pulse (94 mW, $\tau_n=300~\mathrm{ms}$, 800 nm, $\nu_{rep}=80~\mathrm{MHz}$, $B_n=+0.9~\mathrm{G}$), and the final domain after illumination by many trains of low power pulses, as described in the main-text.  Following nucleation, many measurements of the laser induced displacement of the right-hand DW are made. For each measurement, the laser is randomly repositioned within $4~\mathrm{\mu m}$ of the DW and illuminated by a train of ps-pulses (34-mW, $\tau_p=10~\mathrm{ms}, 780~\mathrm{nm}$, $\nu_{rep}=80~\mathrm{MHz}$, $B_{ext}=0$), below the threshold for domain nucleation. Following the application of many ($>50$) pulse-trains, the DW has moved to the right (left) for $\sigma^{\pm}$-polarization and remains relatively unchanged for linear polarization.
(d) Examples of the change in the magnetic domain following excitation with a linearly polarized pulses for different initial positions of the DW with respect to the laser. A cross-section of the difference between the Kerr images taken before and after the laser exposure is plotted against the position relative to the center of the laser spot. Positive (negative) signal indicates a shrinking (growing) domain. In most cases, the final position indicated by the positive gradient, is close to the center of the laser spot, regardless of initial position. (e) Plot of the DW displacement, $\Delta x=x_{f}-x_{i}$ vs initial position $x_{i}$. The effect of the pulse-train is to move the DW to a stationary position $x_0$, that is shifted to the right (left) for $\sigma^{\pm}$-polarizations, respectively. (inset) A positive B-field is defined parallel to the optical-axis.
}\label{fig:fig1}
\end{figure}


To prepare a magnetic domain we use thermally assisted magnetization reversal by laser excitation. First the magnetization is saturated using a negative magnetic field, $B_s=-15~\mathrm{G}$, large compared with the coercive-field, $B_c = 4~\mathrm{G}$. The field is then  ramped to a slightly positive value, $B_{n}=+0.9~\mathrm{G}$. The bar is illuminated for $\tau_{n}=200-400~\mathrm{ms}$ at a high power of 94 mW, at a wavelength of 800 nm to generate a single reversed magnetic domain, as shown in figs. \ref{fig:fig1}(a-c)(i).   The experiments are performed at 92~K, where it is relatively easy to reproducibly nucleate single magnetic domain of a similar size.


Optically assisted magnetization reversal has previously been reported in highly resistive GaMnAs using relatively low power HeNe laser excitation  \cite{Astakhov_prl1,Astakhov_prl2} or a single 80-pJ 100-fs laser pulse \cite{Reid_apl}. There, the polarization independent magnetization reversal was attributed to a reduction in the coercive field due to the photo-carrier related suppression of the DW  pinning potential in material of  low ($<1\%$) Mn-concentration. We attribute the laser induced domain nucleation at small applied reversal fields to thermally assisted magnetization reversal and note a helicity dependent threshold.  The helicity dependence of magnetization reversal has not yet been reported in a magnetic-semiconductor, but has been studied intensively in ferrimetals such as GdFeCo \cite{Stanciu_prl,Vahaplar_prb,Mangin_nmat}.  We now focus on laser induced motion of domain walls at laser powers and exposure times well below the threshold for domain nucleation.


After domain preparation, the sample temperature equalizes to the base temperature of 92K at zero magnetic field. To locate the right-hand DW,
the  laser spot is then positioned outside of the reversed domain and the sample is exposed to a train of 4ps laser pulses, (34 mW, 780 nm, 80MHz repetition frequency) for 10 ms.
To probe a change in DW position, Kerr images before and after the laser illumination are compared. If no change is identified, the laser spot is shifted towards the reversed domain by a step of $0.5~\mathrm{\mu m}$. This procedure is repeated until a first change in DW position is observed.

After identifying the DW location, the laser spot is moved to a randomized position within $4~\mathrm{\mu m}$  of the DW and the sample is again illuminated by a train of ps-pulses. This procedure is repeated until about 30 displacements have been detected or until the entire domain has been  erased by the laser  induced DW motion.

The final differential Kerr images shown in figs. 1(a-c)(ii) are obtained from single Kerr images taken at the final domain configuration and after the domain was erased by a saturation field. In fig. 1(a), where $\sigma^+$-polarized laser pulses are used, the final domain is larger than the initial domain, indicating that the DW moved to the right. By contrast, in fig. \ref{fig:fig1}(c), for $\sigma^-$-excitation, the domain has been completely erased, indicating that the DW moves to the left. In fig. \ref{fig:fig1}(b), the DW position remains relatively unchanged in the case of linear polarization.

Figure \ref{fig:fig1}(d) presents examples of the light induced DW motion with respect to the laser position for linear polarization. Cross-sections of differential Kerr images along the bar identify  DW displacements by the non-zero differential Kerr signals. A positive (negative) signal indicates the right-hand DW moves to the left (right), resulting in a shrinking (growing) domain. The edge of positive gradient, indicating the final DW position is independent of the start position. If the laser spot does not overlap with the DW, no DW motion is observed. In fig. \ref{fig:fig1}(e) the displacement of the DW is plotted against the initial position of the DW with respect to the laser spot. The result is a straight line of gradient -1, indicating that a domain at arbitrary initial position $x_i$ moves to a position where the DW is stationary.  For linear polarization, the DW is attracted to the hot-spot at the center of the laser. A similar observation has recently been reported for DW in CoPt \cite{French_conf}. For $\sigma^{\pm}$  circular polarization, the final position of the DW is shifted to the right (left) with respect to the final position measured for linear polarization.  This demonstrates that the direction of laser induced DW motion depends on the helicity of the laser.


\begin{figure}[ht!]
\begin{center}
\vspace{0.2 cm}
\includegraphics[width=\columnwidth]{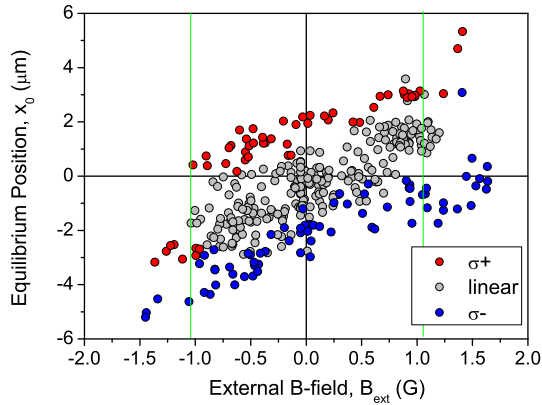}
\vspace{0.2 cm}
\end{center}
\caption{Final DW position $x_0$ versus external magnetic field $B_{ext}$. For small magnetic fields, the equilibrium position varies linearly with magnetic field. The helicity of the laser illumination acts similar to an applied magnetic field that shifts the final position by $x_0 = \pm 2 ~\mathrm{\mu m}$ for $\sigma^{\pm}$ polarization respectively.
}\label{fig:fig2}
\end{figure}



 In the next experiment, a magnetic domain is prepared and the DW is located as before. The laser position is then fixed, and the final position of the DW following excitation by a train of low power pulses is measured as a function of external magnetic field $B_{ext}$. The data is presented in fig. 2.  For a $\sigma^{\pm}$-polarized laser the stationary position $x_0$ is shifted by $\pm 2~\mathrm{\mu m}$ with respect to the case of linear polarization. The gradient is relatively independent of polarization. For small external magnetic fields the final position of the DW can be described by $x_f=x_0+aB_{ext}$, where $x_0$ is the stationary position at $B_{ext}=0$.   We note that for larger external B-fields, ($> 1.5~\mathrm{G}$), the non-illuminated DW can also move as a result of optical excitation of the illuminated DW.


 Above bandgap excitation with $\sigma^{\pm}$ polarization creates photo-carriers with a spin-density $\mathbf{s}\propto \hat{n}=\pm \hat{z}$. Spin-transfer torque mediated by optically generated spin-polarized electrons therefore acts predominantly on the DW where the magnetization rotates into the sample plane. Due to the exchange interaction, the carrier-spin experiences many sub-ps period precessions about the quasi-stationary magnetization vector (period $\sim 10~\mathrm{ns}$) during the 10s' of ps spin-lifetime of the carriers. This results in a time-averaged carrier-spin density along $\mathbf{s}_0\propto \hat{n}\times\hat{m}_{DW}=\pm\hat{y}(\mp\hat{x})$ \cite{Nemec_nphys}, considering  a Neel (Bloch) -type DW with magnetization at the center of the DW along $\hat{x} (\hat{y})$ direction, respectively \cite{deRanieri_nmat}. This kicks the magnetization-vector at the boundary in a direction $\dot{\hat{m}}_{DW}\propto \hat{m}_{DW}\times\hat{s}_0 = \pm\hat{z}$ moving the DW to the right/left, as observed. Hence, optical spin transfer torque can explain the helicity dependence of the direction of DW motion.

 We now argue against the two other candidate mechanisms that could give rise to a helicity dependent shift of the DW position. Firstly, the circular dichroism of the material can lead to a difference in photo-carrier density and temperature across the DW. This would cause the DW to move towards the hot region, as observed for linearly polarized light. In the case of negative saturation magnetic field, the magnetization $\downarrow /\uparrow$ either side of the right-hand DW is $\downarrow\Uparrow\vert\uparrow\Downarrow $, where $\Uparrow / \Downarrow$ indicates the direction of the total angular momentum of the lowest energy heavy-hole state, responsible for the magnetic circular dichroism \cite{Dietl_prb}. In the case of $\sigma^{\pm}$-polarized excitation, a photo-hole of angular momentum $\Downarrow(\Uparrow)$ is added, and the resulting thermal gradient is $hot\vert cold (cold \vert hot)$ causing the DW to move left/right, respectively. This is the opposite to what is observed. Furthermore, the majority of the light, $>95\%$, is absorbed below the 25-nm film of GaMnAs. Therefore, the heating of the sample should be relatively independent of the dichroism.  Hence, the circular dichroism is not the dominant mechanism. Secondly, the laser can generate an effective magnetic field along the optical axis due to the inverse Faraday effect. However, recent studies have shown that compared to the optical spin transfer torque, the inverse Faraday effect is weak in dilute magnetic semiconductors \cite{Nemec_nphys}. These conclusions were made on the basis of spectroscopic measurements showing that the peak in the Kerr rotation was not coincident with that in the measured torque.


\begin{figure}[ht!]
\begin{center}
\includegraphics[width=1\columnwidth]{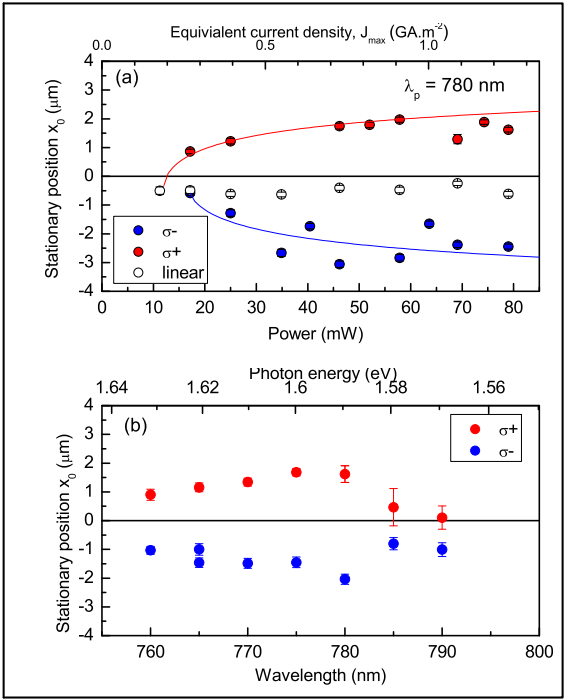}
\end{center}
\caption{Investigation of the stationary position relative to laser position, $x_0$. (a) $x_0$ vs power. The sign of $x_0$ changes with the helicity.  There is a power threshold, followed by saturation.    (b) $x_0$ vs wavelength, ($P=25~\mathrm{mW}$, $\tau_p=5~\mathrm{ms}$). $x_0$ exhibits a broad peak centered at $\sim 775 ~\mathrm{nm}$.
}\label{fig:fig3}
\end{figure}


To further test our understanding, we investigate the power,  and wavelength dependence of the final domain wall position, $x_0$. The results are presented in fig. \ref{fig:fig3}.
Figure 3(a) plots the power dependence of  $x_0$. Assuming the DW moves until reaching a position  where the power density $Pe^{-x^2/w_{eff}^2}$ is below the threshold $P_{th}$ for DW motion, a manual fit to $x_0\approx x_0^{(lin)}\pm w_{eff}\sqrt{ln(\frac{P}{P_{th}})}$ is made.
The  threshold power is $P_{th}=12 (17)~ \mathrm{mW}$ for $\sigma^{\pm}$-polarization respectively \cite{foot_thresh}. This equates to an effective current density (photon flux $\times$ e) of $J_{max}^{th}\approx 0.25~\mathrm{GA.m^{-2}}$. We note that this is similar to the threshold current density measured for electrically driven DW motion in the same wafer \cite{deRanieri_nmat}. The effective width of the Gaussian temperature profile is less than the laser spot-size, $w_{eff}=1.9~\mathrm{\mu m}<w$. This is attributed to a power-threshold that is lower at the hot-spot. Due to the power threshold, we conclude that the DW is moving in a flow regime \cite{deRanieri_nmat} driven by an optical spin transfer torque.

 Figure 3(b) presents the wavelength dependence of $x_0$, which peaks at 785 nm. DW motion is only observed for above band-gap excitation verifying that the DW is driven by photo-generated carriers. The generation of spin may become less effective at higher photon energies due to increased  spin relaxation. In this wavelength regime, the circular dichroism increases monotonically with wavelength \cite{Tesarova_prb}, further ruling out circular dichroism as the dominant source of the helicity dependent term. The helicity dependent direction of DW motion is observed for the experimentally accessible temperatures of 85-98~K.




\begin{figure}[ht!]
\begin{center}
\vspace{0.2 cm}
\includegraphics[width=0.8\columnwidth]{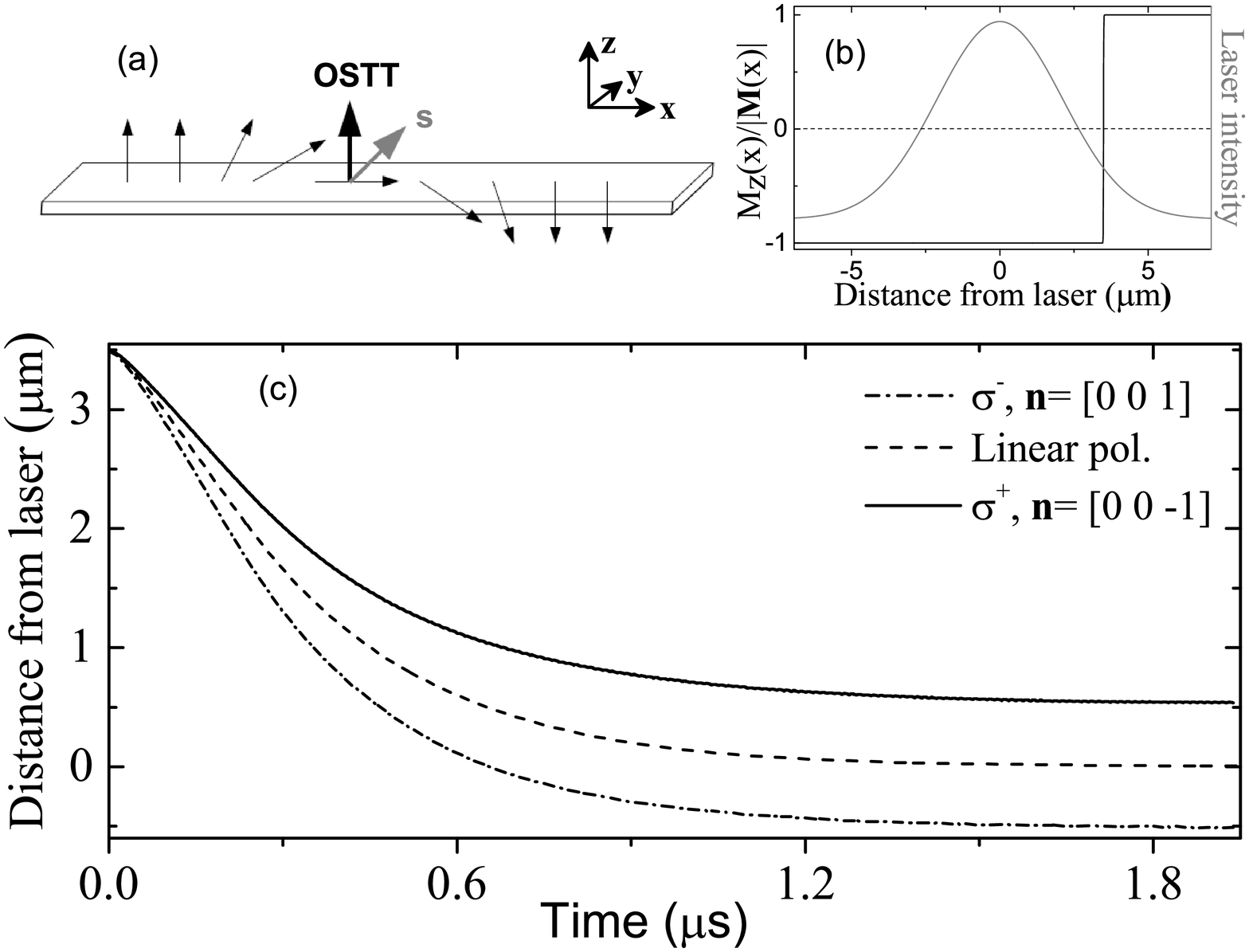}
\vspace{0.2 cm}
\end{center}
\caption{(a) Schematic of magnetization at a N\'eel wall and direction of the optical spin transfer torque. (b) Initial position of DW and and laser spot intensity profile. (c) Subsequent DW motion following application of 80 MHz train of laser pulses. For linear polarization, the DW moves to the center of the laser spot. For circular polarization ($\sigma^{\pm}$) the additional spin transfer torque slows down (speeds up) the DW motion, shifting the final position by $\pm 0.5~\mathrm{\mu m}$. The average speed over the first $0.25~\mathrm{\mu s}$ is about $v_{lin}\approx 5~\mathrm{ms^{-1}}$ or 60 nm/pulse; and $v_-\approx 6~\mathrm{ms^{-1}}$, and $v_+\approx 4~\mathrm{ms^{-1}}$.   }
\label{fig:fig4b}
\end{figure}

To estimate the expected shifts in the DW following laser excitation, simulations using a 1-D micromagnetic Landau-Lifshitz-Bloch (LLB) model \cite{Schiebach_prb}, including the demag field \cite{Newell_jgr}, were performed with parameters given in \cite{LLB}. In the case of linear polarization, a temperature increase proportional to the time-averaged intensity is assumed with an amplitude $\Delta T= 13~K$ \cite{Tesarova_nphoton}, and a base temperature of 92~K. This modifies the magnitude of the equilibrium magnetization $M_{eq}(T)$ \cite{meq} scaling the exchange stiffness and magnetic anisotropy-coefficients of the internal fields, resulting in a free-energy well for the DW. As shown in fig. \ref{fig:fig4b}(b), for the $[1\bar{1}0]$-bar, the N\'eel wall is initially at a position of $+3.5~\mathrm{\mu m}$ from the center of the laser spot. When the temperature profile is switched on, the DW moves to the center of the hot-spot on a microsecond timescale, as shown in fig. \ref{fig:fig4b}(c).


For circular polarization, the optical spin transfer torque is described by an additional effective field in the LLB equation, $\mathbf{H}^{OSTT}_{eff}= \frac{J_{eff}}{\mu_0M_{eq}(T)}\mathbf{s}$, due to the exchange field exerted by the carrier spin density $\mathbf{s}$ on the magnetization. An additional rate equation \cite{Nemec_nphys} is used to describe the time-evolution of the spin: $\mathbf{\dot{s}}=(J_{eff}(T)/m_{eq}\hbar)\mathbf{m}\times\mathbf{s}+R(t)\hat{n}-\mathbf{s}/\tau$, where $R(t)\hat{n}$ describes the spin pumping rate due to  laser excitation, and $\tau\approx 30~\mathrm{ps}$ is the carrier spin-lifetime \cite{Nemec_nphys}. The spin pump rate is treated as a $\nu_{rep}=80~\mathrm{MHz}$ train of square pulses of duration $\tau_{L}=4~\mathrm{ps}$, proportional to the intensity profile of the laser, and an effective pump-rate $e\nu_{rep}\tau_{L}R_{max}=0.8 \mathrm{GA.m^{-2}.\mu m^{-1}}$, which assumes an absorption length of $\sim 1~\mathrm{\mu m}$ in GaAs \cite{abs}. The different polarization cases $\sigma^{\pm}$ are controlled by direction of the carrier-spin $\hat{n}=(0,0,\mp 1)$ respectively. In the simulations shown in fig. \ref{fig:fig4b}(a), the DW moves to an stationary position $x_0$ shifted by $\pm 0.6~\mathrm{\mu m}$ with respect to the center of the laser spot, reproducing the observed helicity dependence of the sign of the shift. Calculations with no temperature gradient and uniform illumination have also been performed. There the DW moves by $\pm 3.5~\mathrm{ms^{-1}}$ in the steady-state. Hence the temperature gradient limits the displacement, and the calculation represents a lower limit on the displacement.  The effect of the OSTT on the DW is illustrated in fig. \ref{fig:fig4b}(a). Initially, the injected carrier-spin is aligned along the z-axis. The precession around the exchange field due to the magnetization is fast compared to the carrier-lifetime $0.4~\mathrm{ps}$ vs $30~\mathrm{ps}$, and the time-averaged carrier spin aligns along the $\hat{y}\propto\hat{n}\times \mathbf{m}$ axis. For the example of a N\'eel wall, the carrier spin applies a torque on the magnetization, kicking the magnetization in the $\pm\hat{z}$ direction. Following the kick, the magnetization precesses around the internal fields moving the DW. For the case of 80-MHz repetition rate considered here, the DW is still moving when the next laser pulse strikes, leading to a steady-state motion where the magnetization precesses around an equilibrium state that is intermediate between a N\'eel and a Bloch wall. We note that similar DW motion is observed and calculated for the Bloch-wall of the [110] bar.

To summarize, we observe shifts in a magnetic DW position following above bandgap excitation with a train of picosecond laser pulses. Two main driving terms are identified and reproduced in micromagnetic simulations. The first helicity dependent term results from spin-polarized photo-carriers exerting a spin transfer torque on the DW.  The second helicity independent term attracts the DW to the laser hot-spot due to  a free energy well resulting from a reduction in the local magnetic moment. Laser manipulation of DW position provides a tool for local rather than global control of DW motion. It provides an experimental  route to investigate DW motion following ultrafast, rather than nanosecond, kicks to the spin transfer torque. By isolating laser induced magnetization reversal from DW propagation, these techniques should provide insights into magnetization reversal.

We acknowledge funding from European Metrology Research Programme within the Joint Research
Project EXL04 (SpinCal); Charles University, Prague, grant no. 1360313; Grant Agency of the Czech Republic under Grant 14-37427G and Hitachi Europe Ltd.

\bibliographystyle{apsrev}

\begin{thebibliography}{1}
\bibitem{foot_thresh}
We attribute the helicity dependence of the threshold to an offset error in the B-field and/or drift in the focusing of the laser spot.
\bibitem{LLB}
Parameters used at zero temperature are $M_{eq}(0)=35.57~\mathrm{kA.m^{-1}}$, $A=0.3~\mathrm{pJ.m^{-1}}$, $K_z=6000~\mathrm{J.m^{-3}}$, $K_x=1150~\mathrm{J.m^{-3}}$, and intrinsic damping $\alpha=0.01$.
$J_{eff}=\frac{5}{2}Jc_{Mn}$, $J=10~\mathrm{meV.nm^3}$, $c_{Mn}=1~\mathrm{nm^{-3}}$, as defined in ref. \cite{deRanieri_nmat}. $A$, $K$ and $J_{eff}$ are scaled by $m^2_{eq}(T)$, where $m_{eq}(T)=M_{eq}/M_{eq}(0)$.
The cell-size is $4\times 4000\times 25~\mathrm{nm^3}$.
\bibitem{meq}
$m_{eq}(T)$ is measured by SQUID, and estimated by a fit to a Langevin function.
\end{thebibliography}

\end{document}